# The effect of gravitational spin-orbit coupling on the circular photon orbit in the Schwarzschild geometry

Zhi-Yong Wang[*], Cai-Dong Xiong, Qi Qiu, Yun-Xiang Wang, Shuang-Jin Shi

*E-mail:   zywang@uestc.edu.cn

*School of Optoelectronic Information, University of Electronic Science and Technology of China, Chengdu 610054, CHINA*

**Abstract**

The $(1,0) \oplus (0,1)$ representation of the group SL(2, C) provides a six-component spinor equivalent to the electromagnetic field tensor. By means of the $(1,0) \oplus (0,1)$ description, one can treat the photon field in curved spacetime via spin connection and the tetrad formalism, which is of great advantage to study the gravitational spin-orbit coupling of photons. Once the gravitational spin-orbit coupling is taken into account, the traditional radius of the circular photon orbit in the Schwarzschild geometry should be replaced with two different radiuses corresponding to the photons with the helicities of $\pm 1$, respectively. Owing to the splitting of energy levels induced by the spin-orbit coupling, photons (from Hawking radiations, say) escaping from a Schwarzschild black hole are partially polarized, provided that their initial velocities possess nonzero tangential components.

**PACS numbers**: 04.62.+v, 04.70.Dy , 41.20.Jb

## 1. Introduction

As we know, spatio-temporal tensors are associated with the Lorentz group, while spinors are associated with its covering group SL(2, C), where the group SL(2, C) yields all representations, including those with half-integral spins, whereas the Lorentz group yields only the representations with integral spins. Therefore, spinors can be used to describe particles of *any* spin, whereas tensors can describe only the particles with integer spins. As a consequence, one can associate every tensor with a spinor, but not vice versa: not all spinors correspond to tensors [1]. In this paper, the $(1,0)\oplus(0,1)$ representation of SL(2, C) provides a six-component spinor equivalent to the electromagnetic field tensor. Historically, several formalisms have been used for fields of any spin *s*, where the 2(2*s* +1) formalism gives the equations which are in some sense on an equal footing with the Dirac equation [2, 3]. Likewise, one can apply the $(1,0)\oplus(0,1)$ description to reformulating Maxwell equations as the so-called Dirac-like equation [4-6].

According to the traditional theory, the circular photon orbit in the Schwarzschild geometry is unstable [7]. In this paper, by means of the $(1,0)\oplus(0,1)$ description, we will analyze the effect of the gravitational spin-orbit coupling on such an orbit. The instability of this orbit does not affect our work, because a temporary circular orbit is enough for us. There are three main reasons to make our work be based on the $(1,0)\oplus(0,1)$ description:

1). As we know, the simplest covariant massless field for helicity $\pm 1$ has the Lorentz transformation type $(1,0)\oplus(0,1)$, and it is impossible to construct a vector field for massless particles of helicity $\pm 1$ [2, 3]. This implies that it is advantageous to study the spin-orbit interaction of the photon field based on the $(1,0)\oplus(0,1)$ description.

2). In Ref. [1] all spinors equivalent to tensors are taken as two-component forms, they are too complicated and abstract to be applicable for some practical issues. In our case, the $(1,0)\oplus(0,1)$ spinor equivalent to the electromagnetic field tensor is a six-component form. Our formalism is more convenient and straightforward, based on which our theory can be developed in a manner being closely parallel to that of the $(1/2,0)\oplus(0,1/2)$ Dirac field.

3). By means of the $(1,0)\oplus(0,1)$ description, one can treat the photon field in curved spacetime via spin connection and the tetrad formalism, which is of great advantage to study the gravitational spin-orbit coupling of photons.

This paper is organized as follows. In Section 2 we present an overview of the $(1,0) \oplus (0,1)$ description. Applying the $(1,0) \oplus (0,1)$ description, in Section 3 we present a self-contained argument about how to treat the photon field in curved spacetime by means of spin connection and the tetrad formalism. Based on the results obtained in Section 3, we study the effect of gravitational spin-orbit coupling on the circular photon orbit in the Schwarzschild geometry.

We work in geometrized units, $\hbar = c = G = 1$, the metric signature is $(-,+,+,+)$, and then the four-dimensional (4D) Minkowski metric tensor is denoted by $\eta_{ab} = \text{diag}(-1,1,1,1)$ ($a,b = 0,1,2,3$). Complex conjugation is denoted by $*$ and hermitian conjugation by $\dagger$.

## 2. The base formalism of the $(1,0) \oplus (0,1)$ description

Let $\partial_\mu = \partial/\partial x^\mu$ with $x^\mu = (t, \boldsymbol{x})$, $\nabla = (\partial_1, \partial_2, \partial_3)$, $\partial_t = \partial_0 = \partial/\partial t$. In Minkowski vacuum, the electromagnetic field intensities $\boldsymbol{E} = (E_1, E_2, E_3)$ and $\boldsymbol{H} = (H_1, H_2, H_3)$ satisfy the Maxwell equations

$$\nabla \times \boldsymbol{H} = \partial \boldsymbol{E}/\partial t, \quad \nabla \times \boldsymbol{E} = -\partial \boldsymbol{H}/\partial t, \tag{1}$$

$$\nabla \cdot \boldsymbol{E} = 0, \quad \nabla \cdot \boldsymbol{H} = 0. \tag{2}$$

In what follows, the *column-matrix forms* of the vectors $\boldsymbol{E}$ and $\boldsymbol{H}$ are also denoted as $\boldsymbol{E}$ and $\boldsymbol{H}$, i.e., $\boldsymbol{E} = \begin{pmatrix} E_1 & E_2 & E_3 \end{pmatrix}^{\text{T}}$, $\boldsymbol{H} = \begin{pmatrix} H_1 & H_2 & H_3 \end{pmatrix}^{\text{T}}$ (the superscript T denotes the matrix transpose, the same below). One can show that the dynamical equations (Eq. (1)) actually contain the transversality conditions (Eq. (2)). In view of the fact that one can associate every tensor with a spinor (but not *vice versa*), let us define a six-component spinor $\psi$ in terms of the electromagnetic field tensor, where the spinor $\psi$ corresponds to the $(1,0) \oplus (0,1)$ representation of SL(2, C) (we will further discuss it later). In terms of the spinor $\psi$ one can rewrite Eq. (1) as the so-called Dirac-like equation,

$$i\beta^\mu \partial_\mu \psi(x) = 0, \text{ or } i\partial_t \psi(x) = \hat{H}\psi(x), \tag{3}$$

where $\beta^\mu = (\beta^0, \boldsymbol{\beta})$, $\boldsymbol{\alpha} = \beta^0 \boldsymbol{\beta}$, $\hat{H} = -i\boldsymbol{\alpha} \cdot \nabla$ represents the Hamiltonian of photons in Minkowski vacuum, and, in the standard representation of the spinor $\psi$, one has

$$\psi = \frac{1}{\sqrt{2}} \begin{pmatrix} \boldsymbol{E} \\ i\boldsymbol{H} \end{pmatrix}, \ \beta^0 = \begin{pmatrix} I_{3\times 3} & 0 \\ 0 & -I_{3\times 3} \end{pmatrix}, \ \boldsymbol{\beta} = \begin{pmatrix} 0 & \boldsymbol{\tau} \\ -\boldsymbol{\tau} & 0 \end{pmatrix}, \ \boldsymbol{\alpha} = \beta^0 \boldsymbol{\beta} = \begin{pmatrix} 0 & \boldsymbol{\tau} \\ \boldsymbol{\tau} & 0 \end{pmatrix}, \tag{4}$$

where $I_{n\times n}$ denotes the $n\times n$ unit matrix ($n = 2, 3,...$), and the matrix vector $\boldsymbol{\tau} = (\tau_1, \tau_2, \tau_3)$ consists of three components:

$$\tau_1 = \begin{pmatrix} 0 & 0 & 0 \\ 0 & 0 & -i \\ 0 & i & 0 \end{pmatrix}, \ \tau_2 = \begin{pmatrix} 0 & 0 & i \\ 0 & 0 & 0 \\ -i & 0 & 0 \end{pmatrix}, \ \tau_3 = \begin{pmatrix} 0 & -i & 0 \\ i & 0 & 0 \\ 0 & 0 & 0 \end{pmatrix}. \tag{5}$$

Let $\hat{\boldsymbol{L}} = \boldsymbol{x} \times (-i\nabla)$, one can easily prove $[\hat{H}, \hat{\boldsymbol{L}} + \boldsymbol{\Sigma}] = 0$, where $\boldsymbol{\Sigma} = I_{2\times 2} \otimes \boldsymbol{\tau}$ satisfying $\boldsymbol{\Sigma} \cdot \boldsymbol{\Sigma} = 2I_{6\times 6}$ is the 3D spin matrix of the photon field in the $(1,0) \oplus (0,1)$ description. In addition to the standard representation mentioned above, one can also define the chiral representation of the spinor $\psi$ via the unitary transformations of $\psi_C = U\psi$ and $\beta_C^\mu = U\beta^\mu U^{-1}$ (and so on), where the subscript C denotes the chiral representation, and the unitary matrix $U$ satisfies

$$U = U^\dagger = U^{-1} = \frac{1}{\sqrt{2}} \begin{pmatrix} I_{3\times 3} & I_{3\times 3} \\ I_{3\times 3} & -I_{3\times 3} \end{pmatrix}. \tag{6}$$

It is easy to show that $\boldsymbol{\Sigma}_C = \boldsymbol{\Sigma}$, $\boldsymbol{\alpha}_C = \beta_C^0 \boldsymbol{\beta}_C$, and

$$\psi_C = \frac{1}{2} \begin{pmatrix} \boldsymbol{E} + i\boldsymbol{H} \\ \boldsymbol{E} - i\boldsymbol{H} \end{pmatrix}, \ \beta_C^0 = \begin{pmatrix} 0 & I_{3\times 3} \\ I_{3\times 3} & 0 \end{pmatrix}, \ \boldsymbol{\beta}_C = \begin{pmatrix} 0 & -\boldsymbol{\tau} \\ \boldsymbol{\tau} & 0 \end{pmatrix}, \ \boldsymbol{\alpha}_C = \begin{pmatrix} \boldsymbol{\tau} & 0 \\ 0 & -\boldsymbol{\tau} \end{pmatrix}. \tag{7}$$

In fact, the chiral and standard representations of the $(1,0) \oplus (0,1)$ field (satisfying the Dirac-like equation) are respectively analogous to the chiral and standard representations of the $(1/2,0) \oplus (0,1/2)$ field (satisfying the Dirac equation) [7], but for the Pauli matrix vector $\boldsymbol{\sigma} = (\sigma_1, \sigma_2, \sigma_3)$ being replaced with the matrix vector $\boldsymbol{\tau} = (\tau_1, \tau_2, \tau_3)$. Moreover, let $\psi^\dagger$ denote the Hermitian adjoint of $\psi$ and $\bar{\psi} = \psi^\dagger \beta^0$, one can show that the quantity $\bar{\psi}\psi$ is a Lorentz scalar, while $\psi^\dagger \psi$ corresponds to the energy density of the photon field, these properties are also similar to those of the Dirac field. The quantization theory of the photon field can easily be developed by means of the $(1,0) \oplus (0,1)$ description.

More generally, under a Lorentz transformation parametrized by an antisymmetric tensor $\omega^{\mu\nu} = -\omega^{\nu\mu}$, $x^\mu \to x'^\mu = \Lambda^\mu{}_\nu x^\nu$, a field quantity $\varphi$ is called a spinor provided that it transforms in the following manner:

$$\varphi(x) \to \varphi'(x') = \exp(-i\omega^{\mu\nu} S_{\mu\nu}/2)\varphi(x) = L(\Lambda)\varphi(x), \tag{8}$$

where $L(\Lambda) = \exp(-i\omega^{\mu\nu} S_{\mu\nu}/2)$ is called the spinor representation of the Lorentz transformation $\Lambda$ (i.e., the representation of SL(2, C)), and the antisymmetric tensor $S_{\mu\nu} = -S_{\nu\mu}$ is the 4D spin tensor of the field $\varphi$. In our case, the six-component spinor $\varphi = \psi$ corresponds to the $(1,0) \oplus (0,1)$ representation of SL(2, C), and

$$S_{lm} = \varepsilon_{lmn}\Sigma^n, \quad S_{0l} = -i\alpha_l, \quad k,l,m = 1,2,3, \tag{9}$$

where the matrices $\Sigma_l$ and $\alpha_l$ are given by Eq. (4) or Eq. (7), $\varepsilon_{klm} = \varepsilon^{klm}$ denotes the totally antisymmetric tensor with $\varepsilon_{123} = 1$.

Now let us present several formulae useful for our future work. Let $a^\mu = (a^0, \boldsymbol{a})$ and $b^\mu = (b^0, \boldsymbol{b})$ be two 4D vectors, but here the column matrix forms of $\boldsymbol{a}$ and $\boldsymbol{b}$ are denoted as $\boldsymbol{a}_M = \begin{pmatrix} a_1 & a_2 & a_3 \end{pmatrix}^T$, $\boldsymbol{b}_M = \begin{pmatrix} b_1 & b_2 & b_3 \end{pmatrix}^T$, respectively. Let $a = |\boldsymbol{a}|$, one can prove that,

$$(\boldsymbol{\tau}\cdot\boldsymbol{a})(\boldsymbol{\tau}\cdot\boldsymbol{b}) = \boldsymbol{a}\cdot\boldsymbol{b} + i\boldsymbol{\tau}\cdot(\boldsymbol{a}\times\boldsymbol{b}) - \boldsymbol{a}_M \boldsymbol{b}_M^T, \tag{10}$$

$$(\beta^\mu a_\mu)(\beta^\nu b_\nu) = -a^\mu b_\mu - i\boldsymbol{\Sigma}\cdot(\boldsymbol{a}\times\boldsymbol{b}) + \boldsymbol{\alpha}\cdot(\boldsymbol{a}b^0 - a^0\boldsymbol{b}) + I_{2\times 2}\otimes \boldsymbol{a}_M\boldsymbol{b}_M^T, \quad (11)$$

$$(\boldsymbol{a}\cdot\boldsymbol{\tau})^{2l+1} = a^{2l}(\boldsymbol{a}\cdot\boldsymbol{\tau}), \quad (\boldsymbol{a}\cdot\boldsymbol{\tau})^{2l+2} = a^{2l}(\boldsymbol{a}\cdot\boldsymbol{\tau})^2, \quad l = 1, 2, \ldots, \quad (12)$$

$$\exp(\pm i\boldsymbol{a}\cdot\boldsymbol{\tau}) = \cos a \pm i\frac{(\boldsymbol{a}\cdot\boldsymbol{\tau})}{a}\sin a + \frac{\boldsymbol{a}_M\boldsymbol{a}_M^T}{a^2}(1-\cos a). \quad (13)$$

Here we have left out the unit matrices $I_{3\times 3}$ and $I_{6\times 6}$ that are equivalent to 1. In fact, using Eqs. (9) and (13) one can prove Eq. (8).

## 3. The $(1,0)\oplus(0,1)$ description in curved spacetime

Based on the $(1,0)\oplus(0,1)$ spinor formalism, we will for the first time treat the photon field in curved spacetime by means of spin connection and the tetrad formalism, which is of great advantage to study the gravitational spin-orbit coupling of photons. In order to make our argument the more explicit as possible, it is helpful to present a self-contained and systematical argument, and then let us start from a brief review of the tetrad formalism.

### 3.1 Brief review of tetrad formalism

It is important to clarify the notations we are going to use in what follows. From now on, small Latin indices $a$, $b$, ... will range from 0 to 3, and will be used to denote tensor indices in the flat tangent space (namely, they are indices labeling tensor representations of the local Lorentz group, raised and lowered by the Minkowski metric). The Greek indices $\mu$, $\nu$, ... also will range from 0 to 3, but they will refer to tensor objects defined on the Riemann manifold (hence they transform covariantly under diffeomorphisms, and are raised and lowered by the Riemann metric).

As a brief review of the tetrad formalism, we will just mention several key points.

To locally characterize the geometry of a four-dimensional Riemann spacetime, one can introduce a tetrad which is a set of axes $\boldsymbol{\eta}_a = \{\boldsymbol{\eta}_0, \boldsymbol{\eta}_1, \boldsymbol{\eta}_2, \boldsymbol{\eta}_3\}$ attached to each point $x^\mu$ of spacetime. We will choose an orthonormal tetrad, where the axes form a locally inertial frame at each point, so that the orthonormal tetrad forms an orthonormal basis in the local Minkowski space tangent to the Riemann spacetime at the given point $x^\mu$, and they are orthonormal with respect to the Minkowski metric $\eta_{ab}$ of the tangent manifold, i.e. they satisfy the condition $\boldsymbol{\eta}_a \cdot \boldsymbol{\eta}_b = \eta_{ab}$.

The vierbein $V_a{}^\mu$ is defined to be the matrix that transforms between the tetrad frame and the coordinate frame (the tetrad index $a$ comes first, then the coordinate index $\mu$): $\boldsymbol{\eta}_a = V_a{}^\mu \boldsymbol{g}_\mu$, where the basis of coordinate tangent vectors $\boldsymbol{g}_\mu = \{\boldsymbol{g}_0, \boldsymbol{g}_1, \boldsymbol{g}_2, \boldsymbol{g}_3\}$ satisfy $\boldsymbol{g}_\mu \cdot \boldsymbol{g}_\nu = g_{\mu\nu}$. The inverse vierbein $V^a{}_\mu$ is defined to be the matrix inverse of the vierbein $V_a{}^\mu$, so that $V^a{}_\mu V_a{}^\nu = \delta^\nu_\mu$, $V^b{}_\mu V_a{}^\mu = \delta^b_a$, it follows that the coordinate metric is $g_{\mu\nu} = \eta_{ab} V^a{}_\mu V^b{}_\nu$.

The tetrads $\boldsymbol{\eta}_a$ refer to axes that transform under local Lorentz transformations, while the coordinate tangent vectors $\boldsymbol{g}_\mu$ refer to axes that transform under general coordinate transformations. The indices on a coordinate vector or tensor were lowered and raised with the coordinate metric $g_{\mu\nu}$ and its inverse $g^{\mu\nu}$, the indices on a tetrad vector or tensor are lowered and raised with the tetrad metric $\eta_{ab}$ and its inverse $\eta^{ab}$.

A 4-vector can be written in a coordinate- and tetrad- independent fashion as an abstract 4-vector $\boldsymbol{A} = \boldsymbol{\eta}_a A^a = \boldsymbol{g}_\mu A^\mu$, its tetrad and coordinate components are related by the vierbein: $A_a = V_a{}^\mu A_\mu$, $A_\mu = V^a{}_\mu A_a$. The scalar product of two 4-vectors may be $\boldsymbol{A}$ and $\boldsymbol{B}$ may be written variously: $\boldsymbol{A} \cdot \boldsymbol{B} = A_a B^a = A_\mu B^\mu$.

A coordinate vector $A^\mu$ (with a Greek index called curved index) does not change under a tetrad transformation (i.e., a local Lorentz (tangent-space) transformation), and is therefore a tetrad scalar; A tetrad vector $A^a$ (with a Latin index called flat index) does not change under a general coordinate transformation, and is therefore a coordinate scalar. The vierbeins transform as general-covariant vectors with respect to their curved index, and as Lorentz vectors with respect to the flat index:

$$V_a^\mu \to V_a'^\mu = \frac{\partial x'^\mu}{\partial x^\nu} \Lambda_a^{\ b} V_b^\nu. \tag{14}$$

Directed derivatives $e_a$ are defined to be the directional derivatives along the axes $\eta_a$:

$$e_a = \eta_a \cdot \partial = \eta_a \cdot g^\mu \partial_\mu = V_a^{\ \mu} \partial_\mu. \tag{15}$$

Obviously, $e_a$ is a tetrad 4-vector. Unlike coordinate derivatives $\partial_\mu = \partial/\partial x^\mu$, directed derivatives $e_a$ do not commute, their commutator is

$$[e_a, e_b] = [V_a^{\ \mu} \partial_\mu, V_b^{\ \nu} \partial_\nu] = [V_a^{\ \nu}(\partial_\nu V_b^{\ \mu}) - V_b^{\ \nu}(\partial_\nu V_a^{\ \mu})]\partial_\mu \equiv C_{ab}^{\ \ c} e_c. \tag{16}$$

Let $C_{abc} = \eta_{dc} C_{ab}^{\ \ d}$, using Eq. (16), $e_c = V_c^{\ \mu} \partial_\mu$ and $V^b_{\ \mu} V_a^{\ \mu} = \delta_a^b$, one has:

$$C_{abc} = \eta_{dc} C_{ab}^{\ \ d} = \eta_{dc} V^d_{\ \mu} [V_a^{\ \nu}(\partial_\nu V_b^{\ \mu}) - V_b^{\ \nu}(\partial_\nu V_a^{\ \mu})]. \tag{17}$$

In a word, the equivalence principle requires local spacetime structure be identified with Minkowski spacetime possessing Lorentz symmetry. In order to relate local Lorentz symmetry to curved spacetime, one need to solder the local (tangent) space to the external (curved) space. The soldering tools are tetrad fields, and one can use the tetrad formalism to treat spinor fields in general relativity. To do so, one can construct a vierbein field $V_a^{\ \mu}(x)$ at every point in spacetime, it is a set of four orthonormal vectors which defines a frame, and because of the equivalence principle this frame can be made inertial at every point. Physical quantities have separate transformation properties in world space (with metric $g_{\mu\nu}$) and tangent space (with metric $\eta_{ab}$). The vierbein field $V_a^{\ \mu}(x)$ is a contravariant vector in world space and a covariant vector in tangent space.

**3.2 Tetrad covariant derivative in Dirac-like equation**

The geometric description of gravity has been developed using the notions of Riemannian metric $g$ and Christoffel connection $\Gamma$, where the curvature of the spacetime manifold, its dynamical evolution, and its interaction with the matter sources has been described in terms of differential equations for the variables $g$ and $\Gamma$. On the other hand, there has been an alternative (but fully equivalent) approach to the description of a Riemannian manifold based on the notions of vierbein $V$ and Lorentz connection (also called spin connection) $\Omega$ [8]. This alternative language is particularly appropriate to embed spinor fields in a curved spacetime. This alternative geometric formalism naturally leads to the formulation of general relativity as a gauge theory for a local Lorentz group, thus putting gravity on the same footing of the other fundamental interactions.

A general-covariant geometric model, adapted to a curved space–time manifold, must be locally Lorentz invariant if referred to the tangent-space manifold described by the vierbein formalism. The generic transformation of the proper orthochronous Lorentz group can be represented by Eq. (8), in terms of the present the notations, it is:

$$L(\Lambda) = \exp(-i\,\omega_{ab} S^{ab}/2). \tag{18}$$

The six generators $S_{ab} = -S_{ba}$ satisfy the Lie algebra of SO(3, 1):

$$i[S_{ab}, S_{cd}] = \eta_{bc} S_{ad} - \eta_{ac} S_{bd} + \eta_{db} S_{ca} - \eta_{da} S_{cb}. \tag{19}$$

In order to restore the symmetry for local transformations with $\omega_{ab} = \omega_{ab}(x)$, we must associate the six generators $S_{ab}$ with six independent gauge vectors, i.e., the Lorentz connection (or spin connection) $\Omega_\mu$,

$$\Omega_\mu = \Omega^{ab}{}_\mu S_{ab}/2 = \Omega_{ab\mu} S^{ab}/2, \quad \Omega^{ab}{}_\mu = -\Omega^{ba}{}_\mu, \tag{20}$$

and introduce a Lorentz covariant derivative defined by:

$$D_\mu = \partial_\mu - i\Omega_\mu = \partial_\mu - i\Omega^{ab}{}_\mu S_{ab}/2 \ . \tag{21}$$

The tetrad-frame formulae look entirely similar to the coordinate-frame formulae, with the replacement of coordinate partial derivatives by directed derivatives, $\partial_\mu = \partial/\partial x^\mu \to e_a$, and the replacement of coordinate-frame connections by tetrad-frame connections. Using Eqs. (15) and (21), one can show that the tetrad-frame spin connection is

$$D_a = V_a{}^\mu D_\mu = e_a - i\Gamma_{bca} S^{bc}/2, \tag{22}$$

where $\Gamma_{bca} = V_a{}^\mu \Omega_{bc\mu}$. Both $\Gamma_{bca}$ and $\Omega^{ab}{}_\mu$ are called the connection coefficients. In terms of $C_{abc} = \eta_{dc} C_{ab}{}^d$ defined by Eqs. (16) and (17), one can express the connection coefficients $\Gamma_{bca}$ as

$$\Gamma_{bca} = -(C_{bca} + C_{cab} - C_{abc})/2 . \tag{23}$$

Using Eqs. (16) and (17) one can calculate $\Gamma_{bca}$ via Eq. (23).

**3.3 Dirac-like equation in Schwarzschild metric**

As for the photon field in the $(1,0) \oplus (0,1)$ description, the infinitesimal generators satisfying the Lie algebra given by Eq. (19) are given by Eq. (9). The Dirac-like equation in flat Minkowski spacetime is invariant under arbitrary global Lorentz transformations. Within the framework of general relativity, the Dirac-like equation in flat Minkowski spacetime should be replaced by the one in a curved spacetime, such that the resulting equation is invariant under local Lorentz transformations. According to the discussions above, using the covariant derivative $D_a$ given by Eq. (22), one can express the Dirac-like equation in a curved spacetime as:

$$i\beta^a D_a \psi(x) = i\beta^a (e_a - i\Gamma_{bca} S^{bc}/2)\psi(x) = 0, \tag{24}$$

where $\beta^a$'s are given by Eq. (4) or Eq. (7), and $S^{ab}$'s are given by Eq. (9). For the moment, the photon field $\psi(x)$ transforms like a scalar with respect to 'world' transformations, but the $(1,0) \oplus (0,1)$ spinor of the group SL(2, C) in tangent space.

From now on we will focus on the special case of the Schwarzschild metric. Outside a Schwarzschild black hole of mass *M*, the standard form of the Schwarzschild metric is

$$ds^2 = -(1-r_s/r)dt^2 + (1-r_s/r)^{-1}dr^2 + r^2(d\theta^2 + \sin^2\theta d\phi^2), \tag{25}$$

where $r_s = 2M$ is the Schwarzschild radius of the black hole. For the convenience of calculating in our framework, we will rewrite Eq. (25) in the isotropic coordinates. To do so, one can introduce a new radial coordinate [9]

$$\rho = (r - r_s/2 + \sqrt{r^2 - r_s r})/2, \text{ or } r = \rho(1 + r_s/4\rho)^2, \tag{26}$$

which implies that $\rho \to r_s/4$ for $r \to r_s$, and the Schwarzschild metric reads:

$$ds^2 = -(1-r_s/4\rho)^2(1+r_s/4\rho)^{-2}dt^2 + (1+r_s/4\rho)^4(d\rho_1^2 + d\rho_2^2 + d\rho_3^2), \tag{27}$$

where the variables $\rho_1$, $\rho_2$ and $\rho_3$ are defined by $\rho_1 = \rho\sin\theta\cos\phi$, $\rho_2 = \rho\sin\theta\sin\phi$, $\rho_3 = \rho\cos\theta$, respectively. Eq. (27) can be written in the following form:

$$ds^2 = -a_0^2 dt^2 + a_1^2 dx_1^2 + a_2^2 dx_2^2 + a_3^2 dx_3^2. \tag{28}$$

Let $x^\mu = (t, \boldsymbol{x})$ with $x^0 = t$, $\boldsymbol{x} = \boldsymbol{\rho} = (\rho_1, \rho_2, \rho_3)$, $\rho = |\boldsymbol{\rho}|$, and

$$a_0 = (1 - r_s/4\rho)(1 + r_s/4\rho)^{-1}, \ a = a_1 = a_2 = a_3 = (1 + r_s/4\rho)^2. \tag{29}$$

Eq. (27) is the so-called isotropic form of the metric in Schwarzschild spacetime, the gradient operator $\nabla = (\partial_1, \partial_2, \partial_3)$ is defined by $\partial_l = \partial/\partial\rho^l$, $l = 1, 2, 3$.

To apply the connection coefficients in an orthonormal basis, let us rewrite Eq. (28) as

$$ds^2 = -(\theta^0)^2 + (\theta^1)^2 + (\theta^2)^2 + (\theta^3)^2, \tag{30}$$

where $\theta^0 = a_0 \mathbf{d}t$, $\theta^l = a_l \mathbf{d}x^l$ ($l = 1, 2, 3$) with $g_{\mu\nu} = \eta_{\mu\nu} = \text{diag}(-1,1,1,1)$ form an orthonormal basis, For the moment, we do not distinguish between tensor indices in the flat tangent space and those in Riemannian spacetime, and denote them by the Greek indices uniformly. The dual basis of $\theta^\mu = a_\mu \mathbf{d}x^\mu$ is

$$e_\mu = a_\mu^{-1} \partial/\partial x^\mu = a_\mu^{-1} \partial_\mu, \quad \mu = 0, 1, 2, 3. \tag{31}$$

In the dual vector space Eqs. (16) and (17) become:

$$[e_\kappa, e_\lambda] = C_{\kappa\lambda}{}^\mu e_\mu, \quad C_{\kappa\lambda\nu} = \eta_{\mu\nu} C_{\kappa\lambda}{}^\mu, \tag{32}$$

Likewise, in the orthonormal basis with with $g_{\mu\nu} = \eta_{\mu\nu} = \text{diag}(-1,1,1,1)$, the connection coefficients given by Eq. (23) can be rewritten as

$$\Gamma_{\kappa\lambda\mu} = -(C_{\kappa\lambda\mu} + C_{\lambda\mu\kappa} - C_{\mu\kappa\lambda})/2. \tag{33}$$

In the Schwarzschild spacetime, the Dirac-like equation Eq. (24) can be rewritten as

$$\mathrm{i}\beta^\mu (e_\mu - \mathrm{i}\Gamma_{\kappa\lambda\mu} S^{\kappa\lambda}/2)\psi(x) = 0. \tag{34}$$

Using Eqs. (28), (31) and (32), one can prove that ($l, m = 1, 2, 3$ and $l \neq m$)

$$C_{0l0} = -C_{l00} = -a_l^{-1} \partial_l \ln a_0, \quad C_{lmm} = -C_{mlm} = -a_l^{-1} \partial_l \ln a_m, \tag{35}$$

with the others vanishing. Using Eqs. (33) and (35), seeing that $S^{\mu\nu} = -S^{\nu\mu}$ and $C_{\kappa\lambda\nu} = -C_{\lambda\kappa\nu}$, one can prove that (see **Appendix A**)

$$\begin{aligned}\beta^\mu \Gamma_{\kappa\lambda\mu} S^{\kappa\lambda}/2 = &-\beta^0 S^{10} a_1^{-1} \partial_1 \ln a_0 + \beta^2 S^{12} a_1^{-1} \partial_1 \ln a_2 + \beta^3 S^{13} a_1^{-1} \partial_1 \ln a_3 \\ &+ \beta^1 S^{21} a_2^{-1} \partial_2 \ln a_1 + \beta^3 S^{23} a_2^{-1} \partial_2 \ln a_3 + \beta^1 S^{31} a_3^{-1} \partial_3 \ln a_1 + \beta^2 S^{32} a_3^{-1} \partial_3 \ln a_2\end{aligned}. \tag{36}$$

Substituting Eqs. (31) and (36) into Eq. (34), one can obtain

$$\begin{aligned}&a_0^{-1}\mathrm{i}\beta^0 \partial_0 \psi + a_1^{-1}[\mathrm{i}\beta^1 \partial_1 - \beta^0 S^{10} \partial_1 \ln a_0 + \beta^2 S^{12} \partial_1 \ln a_2 - \beta^3 S^{31} \partial_1 \ln a_3]\psi \\ &+ a_2^{-1}[\mathrm{i}\beta^2 \partial_2 - \beta^0 S^{20} \partial_2 \ln a_0 + \beta^3 S^{23} \partial_2 \ln a_3 - \beta^1 S^{12} \partial_2 \ln a_1]\psi \\ &+ a_3^{-1}[\mathrm{i}\beta^3 \partial_3 - \beta^0 S^{30} \partial_3 \ln a_0 + \beta^1 S^{31} \partial_3 \ln a_1 - \beta^2 S^{23} \partial_3 \ln a_2]\psi = 0\end{aligned}. \tag{37}$$

Note that Eq. (37) is not only valid for the photon field (the Dirac-like field in our formalism), but also for the massless Dirac field (for the moment $\beta^\mu = \gamma^\mu$ are the Dirac matrices satisfying $\gamma^\mu \gamma^\nu + \gamma^\nu \gamma^\mu = -2\eta^{\mu\nu}$ and $S^{\mu\nu} = i[\gamma^\mu, \gamma^\nu]/4$, see **Appendix A**).

Using Eq. (4) or Eq. (7), $[\tau_l, \tau_m] = i\varepsilon_{lmn}\tau_n$, one can prove that

$$\beta^l \Sigma^m - \beta^m \Sigma^l = i\varepsilon^{lmn}\beta_n, \quad \beta^l \Sigma^m - \Sigma^m \beta^l = i\varepsilon^{lmn}\beta_n. \tag{38}$$

Using Eqs. (9) and (38), $\beta^0 \alpha^l = \beta^l$, one can show that Eq. (37) becomes

$$\begin{aligned}
& a_0^{-1} i\beta^0 \partial_0 \psi + a_1^{-1} i\beta^1 [\partial_1 + (\partial_1 \ln a_0 a_2)]\psi + a_2^{-1} i\beta^2 [\partial_2 + (\partial_2 \ln a_0 a_3)]\psi \\
& + a_3^{-1} i\beta^3 [\partial_3 + (\partial_3 \ln a_0 a_1)]\psi + a_1^{-1} \beta^3 \Sigma^2 [\partial_1 \ln(a_2/a_3)]\psi \\
& + a_2^{-1} \beta^1 \Sigma^3 [\partial_2 \ln(a_3/a_1)]\psi + a_3^{-1} \beta^2 \Sigma^1 [\partial_3 \ln(a_1/a_2)]\psi = 0
\end{aligned} \tag{39}$$

Substituting Eq. (29) into Eq. (39) and using Eq. (38), one has

$$i\beta^0 \eta \partial_0 \psi + i\beta^l (\partial_l + \Pi_l)\psi(x) = 0, \tag{40}$$

where

$$\eta = a_0^{-1} a = (1 - r_s/4\rho)^{-1}(1 + r_s/4\rho)^3, \tag{41}$$

$$\boldsymbol{\Pi} = \nabla \ln \sqrt{\varpi} = \nabla \ln(1 - r_s/4\rho)(1 + r_s/4\rho), \tag{42}$$

$$\varpi = (a_0 a)^2 = (1 - r_s/4\rho)^2 (1 + r_s/4\rho)^2. \tag{43}$$

Let $\psi = \varpi^{-1/2}\varphi$, $\partial_0' = \eta \partial/\partial t$, $\partial_\mu' = (\partial_0', \nabla) = (\eta \partial_0, \partial_l)$, Eq. (40) can be rewritten as

$$i\beta^\mu \partial_\mu' \varphi = 0. \tag{44}$$

According to the tetrad formalism, the metric tensor for Riemannian spacetime transforms like a scalar with respect to Lorentz transformations in tangent space. As a result, $\eta = a_0^{-1} a$ and $\varpi = (a_0 a)^2$ are two Lorentz scalars in the local Minkowski spacetime. To guarantee Eq. (40) be Lorentz covariant in the local Minkowski spacetime, the constraint conditions should be taken as

$$(\nabla + \boldsymbol{\Pi}) \cdot \boldsymbol{E} = 0, \quad (\nabla + \boldsymbol{\Pi}) \cdot \boldsymbol{H} = 0. \tag{45}$$

It follows from $\psi = \varpi^{-1/2}\varphi$ that $\varphi$ is formed by $\boldsymbol{E}' = \varpi^{1/2} \boldsymbol{E}$ and $\boldsymbol{H}' = \varpi^{1/2} \boldsymbol{H}$, in the same way as $\psi$ being formed by $\boldsymbol{E}$ and $\boldsymbol{H}$. In terms of $\boldsymbol{E}'$ and $\boldsymbol{H}'$ Eq. (45) can be

rewritten as

$$\nabla \cdot \boldsymbol{E}' = \nabla \cdot \boldsymbol{H}' = 0. \tag{46}$$

Let

$$\boldsymbol{\Lambda} = \nabla \ln \sqrt{\eta} = \nabla \ln \sqrt{(1 - r_s/4\rho)^{-1}(1 + r_s/4\rho)^3}. \tag{47}$$

Let $f = f(t, \boldsymbol{\rho})$ be a function, using $\partial_0' = \eta \, \partial/\partial t$, $\nabla \eta = 2\eta \boldsymbol{\Lambda}$, one can obtain that

$$\nabla \partial_0' f - \partial_0' \nabla f = 2\boldsymbol{\Lambda} \partial_0' f, \quad \nabla \times \boldsymbol{\Lambda} f = -\boldsymbol{\Lambda} \times \nabla f, \quad \partial_0' \boldsymbol{\Lambda} f = \boldsymbol{\Lambda} \partial_0' f. \tag{48}$$

$$\partial_0 \boldsymbol{\Pi} f = \boldsymbol{\Pi} \partial_0 f, \quad \nabla \times \boldsymbol{\Pi} f = -\boldsymbol{\Pi} \times \nabla f. \tag{49}$$

It is difficult to solve the exact solutions of Eq. (40) or Eq. (44). However, for our purpose, we will just study a special case in next section.

## 4. Gravitational spin-orbit interaction of the photon field

There is no unique vacuum sate in a general spacetime, and then all of our discussions will be presented from the point of view of a distant observer with respect to the Schwarzschild black hole. Many investigations on the trajectories of light in the Schwarzschild metric have been presented [10, 11], and people concluded that there is no stable circular photon orbit in the Schwarzschild geometry [12]. According to the traditional theory, the radius of the circular photon orbit in the Schwarzschild metric given by Eq. (27) is $r = 3r_s/2 = 3M$. On the other hand, to study a splitting of energy levels in some complicated cases, people always resort to the expressions of frequency (or dispersion curves), which is also our research approach in what follows.

To study the effect of gravitational spin-orbit coupling on the circular photon orbit, let us substitute $\sqrt{2}\varphi = \begin{pmatrix} \boldsymbol{E}' & \mathrm{i}\boldsymbol{H}' \end{pmatrix}^\mathrm{T}$ into Eq. (44), using Eq. (46) one has

$$(\boldsymbol{\tau} \cdot \nabla)\boldsymbol{H}' = \mathrm{i}\eta \partial_t \boldsymbol{E}', \quad (\boldsymbol{\tau} \cdot \nabla)\boldsymbol{E}' = -\mathrm{i}\eta \partial_t \boldsymbol{H}'. \tag{50}$$

The column-matrix form of $\nabla = (\partial_1, \partial_2, \partial_3)$ with $\partial_l = \partial/\partial \rho_l$ ($l = 1, 2, 3$) is denoted as $\lambda = \begin{pmatrix} \partial_1 & \partial_2 & \partial_3 \end{pmatrix}^{\mathrm{T}}$, it follows from Eq. (46) that

$$\lambda \lambda^{\mathrm{T}} E' = \lambda \lambda^{\mathrm{T}} H' = 0. \tag{51}$$

Rewriting Eq. (47) as

$$\Lambda = \nabla \ln \sqrt{(1 - r_s/4\rho)^{-1}(1 + r_s/4\rho)^3} = e_\rho \Lambda_\rho, \tag{52}$$

where $e_\rho = \rho/\rho$, and

$$\Lambda_\rho = -\frac{r_s}{8\rho^2}\left[\frac{3}{(1 + r_s/4\rho)} + \frac{1}{(1 - r_s/4\rho)}\right]. \tag{53}$$

Let $f = E', H'$, using $\nabla \eta = 2\eta \Lambda$ and Eqs. (10), (50), (51), one can obtain that

$$\eta^2 \partial_t^2 f = \nabla^2 f - 2\Lambda \cdot \nabla f - 2\mathrm{i}\tau \cdot (\Lambda \times \nabla) f. \tag{54}$$

The last term on the right-hand side of Eq. (54) represents the spin-orbit coupling. Let $f = F \exp(-\mathrm{i}\omega t)$ with $\omega > 0$, Eq. (54) gives

$$\nabla^2 F - 2\Lambda \cdot \nabla F - 2\mathrm{i}\tau \cdot (\Lambda \times \nabla) F + \eta^2 \omega^2 F = 0, \tag{55}$$

Seeing that $\rho_1 = \rho \sin\theta \cos\phi$, $\rho_2 = \rho \sin\theta \sin\phi$, $\rho_3 = \rho \cos\theta$, let us define

$$\begin{cases} \tau_\rho = \tau_1 \sin\theta \cos\phi + \tau_2 \sin\theta \sin\phi + \tau_3 \cos\theta \\ \tau_\theta = \tau_1 \cos\theta \cos\phi + \tau_2 \cos\theta \sin\phi - \tau_3 \sin\theta \\ \tau_\phi = -\tau_1 \sin\phi + \tau_2 \cos\phi \end{cases}. \tag{56}$$

For the moment Eq. (55) can be rewritten as

$$\frac{1}{\rho^2}\frac{\partial}{\partial \rho}(\rho^2 \frac{\partial}{\partial \rho} F) + \frac{1}{\rho^2 \sin\theta}\frac{\partial}{\partial \theta}(\sin\theta \frac{\partial}{\partial \theta} F) + \frac{1}{\rho^2 \sin^2\theta}\frac{\partial^2}{\partial \phi^2} F \\ + \eta^2 \omega^2 F - 2\Lambda_\rho \frac{\partial F}{\partial \rho} - 2\mathrm{i}\Lambda_\rho (\tau_\phi \frac{1}{\rho}\frac{\partial F}{\partial \theta} - \tau_\theta \frac{1}{\rho \sin\theta}\frac{\partial F}{\partial \phi}) = 0 \tag{57}$$

Let $F = \zeta \exp(\mathrm{i} m \phi)$, where $m$ is actually an integer (it plays the role of the quantum number of orbital angular momentum). For circular motion in the equatorial plane one has $\theta = \pi/2$ and $\rho$=constant (i.e., $r$=constant), such that $\partial_\rho \zeta = \partial_\theta \zeta = 0$ and $\tau_\theta = -\tau_3$ (see Eq.

(56)), Eq. (57) becomes

$$(2m\rho^{-1}\Lambda_\rho\tau_3 + \eta^2\omega^2 - m^2\rho^{-2})\zeta = 0. \tag{58}$$

The nonzero-solution conditions of Eq. (58) imply that (where $\eta^2\omega^2 = m^2\rho^{-2}$, related to the longitudinal polarization solution, is discarded)

$$\omega^2 = \frac{m^2}{\eta^2\rho^2} \pm \frac{2m}{\eta^2\rho}\Lambda_\rho. \tag{59}$$

The second term on the right-hand side of Eq. (59) comes from the contribution of the spin-orbit coupling interaction. To gain some insights into Eq. (59), let us take the radius of the circular photon orbit as the traditional one (i.e., $r = 3r_s/2$), approximately, using Eqs. (26), (41) and (53), one has

$$\rho = (2+\sqrt{3})r_s/4, \quad \eta = 12\sqrt{3} - 18, \quad \Lambda_\rho = -2(2-\sqrt{3})/r_s. \tag{60}$$

Substituting Eq. (60) into Eq. (59), one has

$$\omega^2 \equiv \omega_\pm^2 = \frac{4}{27r_s^2}m(m \pm 1) = \frac{4m^2}{27r_s^2} \pm \frac{4m}{27r_s^2}, \tag{61}$$

where $\omega_+ = 4m(m+1)/27r_s^2$ and $\omega_- = 4m(m-1)/27r_s^2$ represent the energies of photons with the spin projections of $\pm 1$, respectively. The second term on the right-hand side of Eq. (61) represents the contribution from the gravitational spin-orbit coupling, without which Eq. (61) would become the result from the traditional theory (see later). In view of the fact that $\omega_+ \neq \omega_-$, the spin-orbit coupling interaction induces a splitting of energy levels.

Actually, once the gravitational spin-orbit coupling is taken into account, the radius of the circular photon orbit is no longer $r = 3r_s/2 = 3M$. To show this, let us pause to review the traditional theory. Let $\sigma$ be an affine parameter along the geodesic $x^\mu(\sigma)$. For photon moving in the equatorial plane of the Schwarzschild geometry (such that $\theta = \pi/2$), one can obtain the 'energy' equation for photon orbits [12]

$$\omega^2(r) = (\frac{\mathrm{d}r}{\mathrm{d}\sigma})^2 + \frac{h^2}{r^2}(1-\frac{r_s}{r}), \qquad (62)$$

In Eq. (62) the quantities $\omega$ and $h$ are two constants, and they are defined as, respectively

$$\omega = (1-\frac{r_s}{r})\frac{\mathrm{d}t}{\mathrm{d}\sigma}, \quad h = r^2\frac{\mathrm{d}\phi}{\mathrm{d}\sigma}. \qquad (63)$$

From the point of view of a distant observer (an observer at rest at infinity), $\omega$ is the total energy of a photon in its orbit, and $h$ equals the orbit angular momentum of the photon. For motion in a circle $r$ is a constant, Eq. (62) becomes

$$\omega^2 = V_{\mathrm{eff}}(r) = \frac{h^2}{r^2}(1-\frac{r_s}{r}), \qquad (64)$$

where $V_{\mathrm{eff}}(r)$ is an effective potential (as compared with the definition of $V_{\mathrm{eff}}(r)$ in Ref. [12], Eq. (64) has a additional multiplication factor $h^2$). Because $V_{\mathrm{eff}}(r)$ has a single maximum at $r = 3r_s/2$ (i.e., the traditional radius of the circular photon orbit), the circular photon orbit is not a stable one. Substituting $r = 3r_s/2$ into Eq. (64) one has

$$\omega^2 = V_{\mathrm{eff}}(r)\big|_{r=3r_s/2} = \frac{4h^2}{27 r_s^2}. \qquad (65)$$

Note that $h$ plays the role of the orbital angular momentum (likewise, in units $\hbar = c = 1$, the quantum number $m$ is also the orbital angular momentum itself). Eq. (59) can be regarded as the quantum-mechanical (semi-classical) counterpart of Eq. (64), with the correspondence of $m \leftrightarrow h$, and the relation between $r$ and $\rho$ is given by Eq. (26).

Now, using Eqs. (41) and (53) we rewrite Eq. (59) as

$$\omega_+^2(\rho) = \frac{m(1-r_s/4\rho)^2}{\rho^2(1+r_s/4\rho)^6}[m-(\frac{3r_s}{4\rho+r_s}+\frac{r_s}{4\rho-r_s})]. \qquad (66)$$

$$\omega_-^2(\rho) = \frac{m(1-r_s/4\rho)^2}{\rho^2(1+r_s/4\rho)^6}[m+(\frac{3r_s}{4\rho+r_s}+\frac{r_s}{4\rho-r_s})]. \qquad (67)$$

Starting from $\mathrm{d}\omega_\pm^2/\mathrm{d}\rho = 0$ one can obtain two quartic equations about $\rho$ (Eq. (26)

implies that $\rho > r_s/4$), each of them gives only one meaningful solution. As an example, let $r_s = 1$, $m = 2$, Eq. (26) implies that the traditional radius of $r = 3r_s/2 = 1.5$ corresponds to the unique value

$$\rho = \rho_0 = (2+\sqrt{3})/4 \approx 0.933. \tag{68}$$

In contrast with Eq. (68), when we take into account the gravitational spin-orbit coupling, substituting $r_s = 1$ and $m = 2$ into $d\omega_+^2/d\rho = 0$, we obtain that

$$64\rho^3 - 112\rho^2 + 40\rho - 3 = 0 \Rightarrow \rho_+ \approx 1.295 > \rho_0 \approx 0.933. \tag{69}$$

Likewise, substituting $r_s = 1$ and $m = 2$ into $d\omega_-^2/d\rho = 0$, one can obtain that

$$128\rho^4 - 32\rho^3 - 80\rho^2 + 22\rho - 1 = 0 \Rightarrow \rho_- \approx 0.783 < \rho_0 \approx 0.933. \tag{70}$$

Therefore, without the gravitational spin-orbit coupling, the radius of the circular photon orbit satisfies $\rho_0 = (2+\sqrt{3})r_s/4$ (corresponding to $r = 3r_s/2$); once the gravitational spin-orbit coupling is taken into account, the radiuses of circular photon orbits satisfy $\rho_+ > \rho_0$ and $\rho_- < \rho_0$, for the photons with the helicities of $\pm 1$, respectively, which is due to the fact that the spin-orbit coupling induces a splitting of energy levels.

Moreover, similar to the fact that $\omega^2 = V_{\text{eff}}(r)$ given by Eq. (64) is an effective potential, $\omega_\pm^2(\rho)$ given by Eqs. (66) and (67) are two effective potentials including the contributions from the spin-orbit coupling. The fact of $\omega_+^2(\rho) \neq \omega_-^2(\rho)$ (or equivalently, $\rho_+ > \rho_0$ and $\rho_- < \rho_0$) implies that the photons with the same initial velocities but different helicities would have different probabilities of escaping from a Schwarzschild black hole (it is not necessary for us to know how much the probabilities are).

## 5. Conclusions

It is advantageous to study the spin-orbit interaction of the photon field based on the $(1,0) \oplus (0,1)$ description, which is due to the fact that, it is impossible to construct a vector field for massless particles of helicity $\pm 1$ [2, 3], while the $(1,0) \oplus (0,1)$ spinor of the group SL(2, C) exhibits simpler Lorentz transformation properties than (1/2, 1/2) four-vector, and it can form the eigenstates of helicity $\pm 1$.

By means of the $(1, 0) \oplus (0, 1)$ description of the photon field, one can treat the photon field in curved spacetime by means of spin connection and the tetrad formalism, which is of great advantage to study the gravitational spin-orbit coupling of photons. We have studied the effect of gravitational spin-orbit coupling on the circular photon orbit outside a Schwarzschild black hole, and have reached the following conclusions: 1) once the gravitational spin-orbit coupling is taken into account, the traditional radius of the circular photon orbit will be replaced with two different radiuses corresponding to the photons with the helicities of $\pm 1$, respectively, which is due to the fact that the gravitational spin-orbit coupling induces a splitting of energy levels; 2) outside a Schwarzschild black hole, when photons (from Hawking radiations, for example) have the initial velocities with nonvanishing tangential components, because of the splitting of energy levels induced by the gravitational spin-orbit coupling, the photons with the same initial velocities (i.e., initial wavenumber vectors) but different helicities would have different probabilities of escaping from the Schwarzschild black hole (it is not necessary for us to know how much the probabilities are), which implies that the photons escaping from the Schwarzschild black hole are partially polarized, rather than completely disorder.

**Acknowledgments**

The work is supported by the National Natrual Science Foundation of China (No. 61271030) and the National Natrual Science Foundation of China (No. 61308041).

## Appendix A Derivations of Eqs. (35)-(37)

In the orthonormal basis with $g_{\mu\nu} = \eta_{\mu\nu} = \text{diag}(-1,1,1,1)$, $\hbar = c = G = 1$, the dual basis of $\theta^\mu = a_\mu \mathbf{d}x^\mu$ is given by Eq. (31), i.e.,

$$e_0 = a_0^{-1} \partial/\partial t = a_0^{-1}\partial_0, \quad e_l = a_l^{-1} \partial/\partial x^l = a_l^{-1}\partial_l, \quad l = 1,2,3. \tag{a1}$$

Using $\eta_{\mu\nu} C_{\kappa\lambda}{}^\mu = C_{\kappa\lambda\nu}$, $C_{\kappa\lambda}{}^0 = -C_{\kappa\lambda 0}$, $C_{\kappa\lambda}{}^l = C_{\kappa\lambda l}$, it follows from $[e_\kappa, e_\lambda] = C_{\kappa\lambda}{}^\mu e_\mu$ that ($l, m, n = 1, 2, 3$):

(1). Owing to $[e_\lambda, e_\lambda] = 0$, one has

$$C_{\lambda\lambda\mu} = 0, \quad \mu, \lambda = 0,1,2,3. \tag{a2}$$

(2). Owing to $[e_0, e_l] = -[e_l, e_0] = C_{0l}{}^\mu e_\mu = -a_0^{-1}(\partial_0 \ln a_l) e_l + a_l^{-1}(\partial_l \ln a_0) e_0$, one has

$$C_{0ll} = -C_{l0l} = -a_0^{-1}(\partial_0 \ln a_l) = 0, \quad C_{0l0} = -C_{l00} = -a_l^{-1}(\partial_l \ln a_0). \tag{a3}$$

(3). For $l \neq m$, one has $[e_l, e_m] = -[e_m, e_l] = C_{lm}{}^\mu e_\mu = -a_l^{-1}\partial_l (\ln a_m) e_m + a_m^{-1}(\partial_m \ln a_l) e_l$, i.e.,

$C_{lmm} = -C_{mlm} = -a_l^{-1}\partial_l(\ln a_m)$, $C_{lml} = -C_{mll} = a_m^{-1}(\partial_m \ln a_l)$, they are equivalent, then

$$C_{lmm} = -C_{mlm} = -a_l^{-1}\partial_l(\ln a_m). \tag{a4}$$

(4). The cases (2) and (3) also imply that

$$C_{0lm} = -C_{l0m} = 0, \quad C_{lmn} = -C_{mln} = 0, \quad l \neq m \neq n. \tag{a5}$$

The statements (1)-(4) exhaust all cases, they together give Eq. (35).

Using Eq. (33), one has $\beta^\mu \Gamma_{\kappa\lambda\mu} S^{\kappa\lambda}/2 = -\beta^\mu (C_{\kappa\lambda\mu} + C_{\lambda\mu\kappa} - C_{\mu\kappa\lambda}) S^{\kappa\lambda}/2$, that is

$$\begin{aligned}\beta^\mu \Gamma_{\kappa\lambda\mu} S^{\kappa\lambda}/2 &= -\beta^0 (C_{\kappa\lambda 0} + C_{\lambda 0\kappa} - C_{0\kappa\lambda}) S^{\kappa\lambda}/2 - \beta^1(C_{\kappa\lambda 1} + C_{\lambda 1\kappa} - C_{1\kappa\lambda})S^{\kappa\lambda}/2 \\ &\quad - \beta^2 (C_{\kappa\lambda 2} + C_{\lambda 2\kappa} - C_{2\kappa\lambda})S^{\kappa\lambda}/2 - \beta^3 (C_{\kappa\lambda 3} + C_{\lambda 3\kappa} - C_{3\kappa\lambda})S^{\kappa\lambda}/2\end{aligned}. \tag{a6}$$

Using Eq. (35) and $S^{\kappa\lambda} = -S^{\lambda\kappa}$, one has

$$\begin{aligned}&-\beta^0 (C_{\kappa\lambda 0} + C_{\lambda 0\kappa} - C_{0\kappa\lambda})S^{\kappa\lambda}/2 \\ &= -\beta^0 (C_{l00} + C_{00l} - C_{0l0})S^{l0}/2 - \beta^0 (C_{0l0} + C_{l00} - C_{00l})S^{0l}/2, \\ &= -\beta^0 S^{l0} a_l^{-1} \partial_l \ln a_0\end{aligned} \tag{a7}$$

$$-\beta^1(C_{\kappa\lambda 1}+C_{\lambda 1\kappa}-C_{1\kappa\lambda})S^{\kappa\lambda}/2$$
$$=-\beta^1(C_{1n1}+C_{n11})S^{1n}/2-\beta^1(C_{n11}-C_{1n1})S^{n1}/2 \quad , \tag{a8}$$
$$=\beta^1 S^{n1}a_n^{-1}\partial_n \ln a_1 = \beta^1 S^{21}a_2^{-1}\partial_2 \ln a_1 + \beta^1 S^{31}a_3^{-1}\partial_3 \ln a_1$$

$$-\beta^2(C_{\kappa\lambda 2}+C_{\lambda 2\kappa}-C_{2\kappa\lambda})S^{\kappa\lambda}/2$$
$$=-\beta^2(C_{2n2}+C_{n22})S^{2n}/2-\beta^2(C_{n22}-C_{2n2})S^{n2}/2 \quad , \tag{a9}$$
$$=\beta^2 S^{n2}a_n^{-1}\partial_n \ln a_2 = \beta^2 S^{12}a_1^{-1}\partial_1 \ln a_2 + \beta^2 S^{32}a_3^{-1}\partial_3 \ln a_2$$

$$-\beta^3(C_{\kappa\lambda 3}+C_{\lambda 3\kappa}-C_{3\kappa\lambda})S^{\kappa\lambda}/2$$
$$=-\beta^3(C_{3n3}+C_{n33})S^{3n}/2-\beta^3(C_{n33}-C_{3n3})S^{n3}/2 \quad . \tag{a10}$$
$$=\beta^3 S^{n3}a_n^{-1}\partial_n \ln a_3 = \beta^3 S^{13}a_1^{-1}\partial_1 \ln a_3 + \beta^3 S^{23}a_2^{-1}\partial_2 \ln a_3$$

Substituting (a7)-(a10) into (a6), one can obtain Eq. (36), i.e.,

$$\beta^\mu \Gamma_{\kappa\lambda\mu}S^{\kappa\lambda}/2 = -\beta^0 S^{l0}a_l^{-1}\partial_l \ln a_0 + \beta^2 S^{12}a_1^{-1}\partial_1 \ln a_2 + \beta^3 S^{13}a_1^{-1}\partial_1 \ln a_3$$
$$+\beta^1 S^{21}a_2^{-1}\partial_2 \ln a_1 + \beta^3 S^{23}a_2^{-1}\partial_2 \ln a_3 + \beta^1 S^{31}a_3^{-1}\partial_3 \ln a_1 + \beta^2 S^{32}a_3^{-1}\partial_3 \ln a_2 \tag{a11}$$

Consider that

$$i\beta^\mu e_\mu \psi = a_0^{-1}i\beta^0 \partial_0 \psi + a_1^{-1}i\beta^1 \partial_1 \psi + a_2^{-1}i\beta^2 \partial_2 \psi + a_3^{-1}i\beta^3 \partial_3 \psi \quad , \tag{a12}$$

substituting (a11) and (a12) into $i\beta^\mu e_\mu \psi(x)+(\beta^\mu \Gamma_{\kappa\lambda\mu}S^{\kappa\lambda}/2)\psi(x)=0$, and using $S^{\kappa\lambda}=-S^{\lambda\kappa}$, one can obtian Eq. (37), i.e.,

$$a_0^{-1}i\beta^0 \partial_0 \psi + a_1^{-1}[i\beta^1 \partial_1 - \beta^0 S^{10}\partial_1 \ln a_0 + \beta^2 S^{12}\partial_1 \ln a_2 - \beta^3 S^{31}\partial_1 \ln a_3]\psi$$
$$+a_2^{-1}[i\beta^2 \partial_2 - \beta^0 S^{20}\partial_2 \ln a_0 + \beta^3 S^{23}\partial_2 \ln a_3 - \beta^1 S^{12}\partial_2 \ln a_1]\psi \tag{a13}$$
$$+a_3^{-1}[i\beta^3 \partial_3 - \beta^0 S^{30}\partial_3 \ln a_0 + \beta^1 S^{31}\partial_3 \ln a_1 - \beta^2 S^{23}\partial_3 \ln a_2]\psi = 0$$

Note that Eq. (a13) (i.e., Eq. (37)) is valid for both the massless Dirac field and the photon field (i.e., the Dirac-like field). For example, let us discuss Eq. (a13) as follows:

1). As $\psi$ represents the massless Dirac field, one has $\beta^\mu = \gamma^\mu$, where $\gamma^\mu$'s are the Dirac matrices satisfying $\gamma^\mu \gamma^\nu + \gamma^\nu \gamma^\mu = -2\eta^{\mu\nu}$ and $S^{\mu\nu}=i[\gamma^\mu,\gamma^\nu]/4$, the matrix vectors of $\boldsymbol{\alpha}$ and $\boldsymbol{\Sigma}$ are defined by $\boldsymbol{\alpha}=\gamma^0\boldsymbol{\gamma}$ and $\boldsymbol{\Sigma}=I_{2\times 2}\otimes\boldsymbol{\sigma}$ ($\boldsymbol{\sigma}$ is the Pauli matrix vector). Using $\gamma^0\alpha^l = \gamma^l$, $S^{lm}=\varepsilon^{lmn}\Sigma_n/2$, $S^{l0}=-i\alpha^l/2$, and $\gamma^l\Sigma^m = -\gamma^m\Sigma^l = i\varepsilon^{lmn}\gamma_n$, Eq. (a13) becomes

$$a_0^{-1}\mathrm{i}\gamma^0\partial_0\psi + a_1^{-1}[\mathrm{i}\gamma^1\partial_1 + \mathrm{i}\frac{1}{2}\gamma^1\partial_1\ln a_0 + \frac{1}{2}\gamma^2\Sigma^3\partial_1\ln a_2 - \frac{1}{2}\gamma^3\Sigma^2\partial_1\ln a_3]\psi$$

$$+a_2^{-1}[\mathrm{i}\gamma^2\partial_2 + \mathrm{i}\frac{1}{2}\gamma^2\partial_2\ln a_0 + \frac{1}{2}\gamma^3\Sigma^1\partial_2\ln a_3 - \frac{1}{2}\gamma^1\Sigma^3\partial_2\ln a_1]\psi \quad ,$$

$$+a_3^{-1}[\mathrm{i}\gamma^3\partial_3 + \mathrm{i}\frac{1}{2}\gamma^3\partial_3\ln a_0 + \frac{1}{2}\gamma^1\Sigma^2\partial_3\ln a_1 - \frac{1}{2}\gamma^2\Sigma^1\partial_3\ln a_2]\psi = 0$$

which can be rewritten as

$$a_0^{-1}\mathrm{i}\gamma^0\partial_0\psi + a_1^{-1}\mathrm{i}\gamma^1[\partial_1 + (\partial_1\ln\sqrt{a_0 a_2 a_3})]\psi$$
$$+a_2^{-1}\mathrm{i}\gamma^2[\partial_2 + (\partial_2\ln\sqrt{a_0 a_1 a_3})]\psi + a_3^{-1}\mathrm{i}\gamma^3[\partial_3 + (\partial_3\ln\sqrt{a_0 a_1 a_2})]\psi = 0 \quad . \tag{a14}$$

2). As $\psi$ represents the photon field (the Dirac-like field), $\beta^\mu$'s are defined by Eq. (4) or Eq. (7), $\beta^0\alpha^l = \beta^l$, for the moment using Eqs. (9) and (38), one can obtain Eq. (39), that is (for the convenience of comparing with Eq. (a14)),

$$a_0^{-1}\mathrm{i}\beta^0\partial_0\psi + a_1^{-1}\mathrm{i}\beta^1[\partial_1 + (\partial_1\ln a_0 a_2)]\psi + a_2^{-1}\mathrm{i}\beta^2[\partial_2 + (\partial_2\ln a_0 a_3)]\psi$$
$$+a_3^{-1}\mathrm{i}\beta^3[\partial_3 + (\partial_3\ln a_0 a_1)]\psi + a_1^{-1}\beta^3\Sigma^2[\partial_1\ln(a_2/a_3)]\psi \quad . \tag{a15}$$
$$+a_2^{-1}\beta^1\Sigma^3[\partial_2\ln(a_3/a_1)]\psi + a_3^{-1}\beta^2\Sigma^1[\partial_3\ln(a_1/a_2)]\psi = 0$$

In fact, starting from Eq. (a14) and in an analogous manner, one can also study the gravitational spin-orbit coupling interaction of the massless Dirac field.